\documentclass[sigconf]{acmart}
\acmConference[ISSTA 2024]{ACM SIGSOFT International Symposium on Software Testing and Analysis}{16-20 September, 2024}{Vienna, Austria}

\usepackage{adjustbox}
\usepackage{algorithm}
\usepackage{algorithmicx}
\usepackage{algpseudocode}
\usepackage{amsmath,amsfonts}
\usepackage{array}
\usepackage{balance}
\usepackage{booktabs}
\usepackage[skip=1pt,labelfont=bf]{caption}
\usepackage{calc}
\usepackage{calligra}
\usepackage{color}
\usepackage{colortbl}
\usepackage{courier}
\usepackage{csvsimple}
\usepackage{enumitem}
\usepackage{fancybox}
\usepackage{fontenc}
\usepackage{graphicx}
\usepackage{listings}
\usepackage{longtable}
\usepackage{lscape}
\usepackage{makecell}
\usepackage{marvosym}
\usepackage{moreverb}
\usepackage{multicol}
\usepackage{multirow}
\usepackage{pifont}
\usepackage{pgfplots}
\usepackage{rotating}
\usepackage{setspace}
\usepackage{siunitx}
\usepackage{soul}
\usepackage{subcaption}
\usepackage{tablefootnote}
\usepackage[most]{tcolorbox}
\usepackage{threeparttable}
\usepackage{tikz}
\usepackage[normalem]{ulem}
\usepackage{url}
\usepackage{wasysym}
\usepackage{xspace}

\usepgfplotslibrary{statistics}
\usetikzlibrary{matrix, shapes.geometric, arrows}
\usetikzlibrary{shapes, arrows, positioning}

\algnewcommand\algorithmicforeach{\textbf{for each}}
\algdef{S}[FOR]{ForEach}[1]{\algorithmicforeach\ #1\ \algorithmicdo}

\newcolumntype{L}[1]{>{\raggedright\let\newline\\\arraybackslash\hspace{0pt}}m{#1}}
\newcolumntype{C}[1]{>{\centering\let\newline\\\arraybackslash\hspace{0pt}}m{#1}}
\newcolumntype{R}[1]{>{\raggedleft\let\newline\\\arraybackslash\hspace{0pt}}m{#1}}

\definecolor{codegreen}{rgb}{0,0.6,0}
\definecolor{codered}{rgb}{1,0,0}
\definecolor{codegray}{rgb}{0.5,0.5,0.5}
\definecolor{codepurple}{rgb}{0.58,0,0.82}
\definecolor{backcolour}{rgb}{0.95,0.95,0.92}
\definecolor{lightgray}{gray}{0.9}

\newboolean{showcomments}
\setboolean{showcomments}{true}
\ifthenelse{\boolean{showcomments}}
 { \newcommand{\mynote}[2]{
      \fbox{\bfseries\sffamily\scriptsize#1}
        {\small$\blacktriangleright$\textsf{\emph{#2}}$\blacktriangleleft$}}}
        { \newcommand{\mynote}[2]{}}
        
\definecolor{DarkOrange}{rgb}{0.8,0.3,0.0}
\definecolor{DarkCyan}{rgb}{0.0, 0.55, 0.55}
\definecolor{DarkCyel}{rgb}{1.0, 0.49, 0.0}
\definecolor{yellow-green}{rgb}{0.6, 0.8, 0.2}

\newcolumntype{?}{!{\vrule width 1pt}}

\newcommand{\etal}{\emph{et~al.}\xspace}
\newcommand{\dataset}{\textsc{TutorCode}\xspace}
\newcommand{\datasetextend}{\textsc{TutorCodePlus}\xspace}
\newcommand{\tool}{\textsc{Cref}\xspace}
\newcommand{\othertool}{\textsc{MultiRegenerate}\xspace}
\newcommand*{\ie}{i.e., }

\newcommand{\find}[1]{
\begin{tcolorbox}[leftrule=1mm,toprule=0mm,bottomrule=0mm,left=1pt,right=2pt,top=2pt,bottom=2pt]
\em #1
\end{tcolorbox}
}

\lstdefinelanguage{mymarkdown}{
    morekeywords={*,\#, \#\#, \#\#\#},
    sensitive=false,
    morecomment=[l]{//},
    morestring=[b]",
    commentstyle=\color{codegreen},
    keywordstyle=\color{magenta},
    numberstyle=\tiny\color{codegray},
    stringstyle=\color{codepurple},
    basicstyle=\tiny,
    breakatwhitespace=false,         
    breaklines=true,
    breakindent=0pt,
    keepspaces=true,                 
    numbers=left,                    
    numbersep=5pt,                  
    showspaces=false,                
    showstringspaces=false,
    showtabs=false,                  
    tabsize=2,
}

\lstdefinestyle{mystyle}{
    commentstyle=\color{codegreen},
    keywordstyle=\color{magenta},
    numberstyle=\small\color{black},
    stringstyle=\color{codepurple},
    basicstyle=\scriptsize\ttfamily,
    breakatwhitespace=false,
    breaklines=true,
    captionpos=b,
    keepspaces=true,
    showspaces=false,
    showstringspaces=false,
    showtabs=false,
    tabsize=2
}

\lstdefinelanguage{diff}{
  morecomment=[f][\color{blue}]{@@},     
  morecomment=[f][\color{red}]-,         
  morecomment=[f][\color{codegreen}]+,       
  morecomment=[f][\color{red}]{---}, 
  morecomment=[f][\color{codegreen}]{+++},
  numberstyle=\tiny\color{codegray},
  numbers=left,                    
  numbersep=5pt,         
}

\setlist{noitemsep} 

\definecolor{darkpastelred}{rgb}{0.76, 0.23, 0.13}
\definecolor{ao(english)}{rgb}{0.0, 0.5, 0.0}

\definecolor{darkpastelred}{rgb}{0.76, 0.23, 0.13}
\definecolor{ao(english)}{rgb}{0.0, 0.5, 0.0}

\newboolean{useblue}
\setboolean{useblue}{false} 

\newcommand{\maybeblue}[1]{%
    \ifthenelse{\boolean{useblue}}%
    {\textcolor{blue}{#1}}%
    {#1}%
}

\AtBeginDocument{%
  }

\setcopyright{none}
\begin{document}

\title{\tool: An LLM-based Conversational Software Repair Framework for Programming Tutors}


\author{Boyang Yang}
\authornote{Co-first authors who contributed equally to this work.}
\authornote{Also affiliated with Jisuan Institute of Technology, Beijing JudaoYouda Network Technology Co. Ltd.}
\affiliation{%
  \institution{School of Information Science and Engineering, Yanshan University}
  \country{China}
}
\email{yangboyang@jisuanke.com}

\author{Haoye Tian}
\authornotemark[1]
\affiliation{%
  \institution{CIS, University of Melbourne}
  \country{Australia}
}
\email{haoye.tian@unimelb.edu.au}

\author{Weiguo Pian}
\affiliation{%
  \institution{SnT, University of Luxembourg}
  \country{Luxembourg}
}
\email{weiguo.pian@uni.lu}

\author{Haoran Yu}
\affiliation{%
  \institution{Jisuan Institute of Technology, Beijing JudaoYouda Network Technology Co. Ltd.}
  \country{China}
}
\email{yuhaoran@jisuanke.com}

\author{Haitao Wang}
\affiliation{%
  \institution{Jisuan Institute of Technology, Beijing JudaoYouda Network Technology Co. Ltd.}
  \country{China}
}
\email{wanghaitao@jisuanke.com}

\author{Jacques Klein}
\affiliation{%
  \institution{SnT, University of Luxembourg}
  \country{Luxembourg}
}
\email{jacques.klein@uni.lu}

\author{Tegawendé F. Bissyandé}
\affiliation{%
  \institution{SnT, University of Luxembourg}
  \country{Luxembourg}
}
\email{tegawende.bissyande@uni.lu}

\author{Shunfu Jin}
\authornote{Corresponding author.}
\affiliation{%
  \institution{School of Information Science and Engineering, Yanshan University}
  \country{China}
}
\email{jsf@ysu.edu.cn}

\renewcommand{\shortauthors}{Yang and Tian et al}

\begin{abstract}
Program repair techniques offer cost-saving benefits for debugging within software development and programming education scenarios. With the proven effectiveness of \textbf{L}arge \textbf{L}anguage \textbf{M}odels (LLMs) in code-related tasks, researchers have explored their potential for program repair. However, it is crucial to recognize that existing repair benchmarks may have influenced LLM training data, potentially causing data leakage. To evaluate LLMs' realistic repair capabilities, \ding{182} we introduce an extensive, non-crawled benchmark, referred to as \dataset, comprising 1,239 C++ defect codes and associated information such as tutor guidance, solution description, failing test cases, and the corrected code. Our work assesses the repair performance of 12 LLMs on \dataset, measuring repair correctness (TOP-5 and AVG-5) and patch precision (RPSR). \ding{183} We then provide a comprehensive investigation into which types of extra information can help LLMs improve their performance in repairing defects.
Among these types, tutor guidance was found to be the most effective information in enhancing LLM repair capabilities. To fully harness LLMs' conversational capabilities and the benefits of augmented information, \ding{184} we introduce a novel conversational semi-automatic repair framework \tool assisting human programming tutor. It demonstrates a remarkable AVG-5 improvement of 17.2\%-24.6\% compared to the baseline, achieving an impressive AVG-5 of 76.6\% when utilizing GPT-4. These results highlight the potential for enhancing LLMs' repair capabilities through interactions with tutors and historical conversations involving incorrect responses. The successful application of \tool in a real-world educational setting demonstrates its effectiveness in reducing tutors' workload and improving students' learning experience, while also showcasing its promise for facilitating other software engineering tasks, such as code review.
\end{abstract}

\maketitle

\section{Introduction}

In the code-related scenarios, such as programming education, providing efficient and precise automated feedback, especially auto-generated program repairs, is essential for effectively guiding a large number of students and reducing the workload of human tutors~\cite{ahmed2022verifix,kim2023automated,zhang2022automated}. Data from a company's online programming education platform show that 54.5\% of students need debugging assistance while completing programming tasks. Each tutor spends an average of 26.7 minutes resolving a single issue, which leads to high labor costs for the company. Additionally, this extensive resolution time adversely affects students' completion times and overall learning experience.
Program repair techniques, vital in both programming education and software development, significantly reduce the manual labor involved in debugging~\cite{gao2022program, tian2022change, goues2019automated, tian2020evaluating, le2021automatic, le2016history}.
Nowadays, machine learning-based approaches have gained prominence in the field of program repair research. These techniques predominantly rely on \textbf{N}eural \textbf{M}achine \textbf{T}ranslation~(NMT) to correct code~\cite{pian2023metatptrans, ye2022neural, qiu2021deep,ni2022best,parasaram2023rete,jiang2023knod,lin2022context}. Most recently, 
Jiang~\etal~\cite{jiang2023knod} introduced KNOD, an NMT-based program repair method that employs a three-stage tree decoder to capture code structure and perform domain knowledge distillation; Parasaram~\etal~\cite{parasaram2023rete} proposed Rete, a learning-based program repair method that learns namespace representations to navigate the search space of patches, thereby enhancing program repair; Meng~\etal~\cite{meng2023template} introduced TENURE, an innovative template-based neural program repair approach that combines template-based and NMT-based methods, demonstrating superior performance compared to other machine learning-based approaches when evaluated on the Defects4J dataset.

Code-targeted pre-trained LLMs, such as CodeGen~\cite{nijkamp2022codegen}, Incoder \cite{fried2022incoder}, and StarCoder~\cite{li2023starcoder}, have made significant strides in advancing the field of machine learning-based program repair. These LLMs have been instrumental in generative program repair through various paradigms. This includes zero-shot approaches, which utilize the original incorrect code either with~\cite{kolak2022patch, prenner2022can,fan2023automated} or without accompanying instructions~\cite{fu2022vulrepair}, and few-shot approaches that incorporate a small set of patch examples~\cite{phung2023generating}. 
With the expansion of training datasets to include large-scale code-related data crawled from the internet~\cite{openai2023gpt4,brown2020gpt3,touvron2023llama}, there has been a noticeable enhancement in LLMs' program repair capabilities~\cite{li2023starcoder}. 
However, the inherent stochastic and opaque nature of LLMs makes the generated patches unreliable~\cite{azaria2023chatgpt}. Yet, many of the current LLM-based repair techniques typically treat the repair task as an automated procedure, often overlooking the collaborative and interactive aspects inherent in programming~\cite{zhang2023steam}. To harness the conversational capabilities of LLMs and address the unpredictability of generated code, there is a growing need for engaging in interactive program repair~\cite{ge2023llm,xia2023conversation}. For instance, Gao~\etal~\cite{gao2020interactive} introduced interactive program repair where developers are involved in reviewing and selecting automatically generated patches. Building on the success of recent LLM-as-a-service deployments, Sobania~\etal~\cite{sobania2023analysis} enhanced the repair performance of ChatGPT by incorporating human-authored hints. They have validated their approach on QuixBugs, based on a corpus of 40 bugs. From a different perspective, Xia~\etal~\cite{xia2023conversation} proposed ChatRepair, where 
 when an LLM-generated patch fails to pass a test case, a new prompt is constructed by combining the invalid patch with the failing test case information, towards generating the next prompt. This process is executed in at most three turns of dialogs, and has been validated on 337 bugs from QuixBugs and  Defects4J. Considering the importance of utilizing ChatGPT judiciously and with expertise, Azaria~\etal~\cite{azaria2023chatgpt} proposed an LLM-based repair strategy designed for experts who are well-versed in the respective domains. Unfortunately, they did not perform any performance evaluation.

Overall, however, interactive program repair using LLMs still faces several limitations:

\begin{itemize}[leftmargin=*]
    \item[\ding{172}] {Data Leakage Concerns}: The effectiveness of LLMs in interactive program repair often relies on the scale of code-related data collected from the internet, which can be associated with {\bf data leakage} issues. Indeed, the benchmarks used for evaluation may have been part of the training data for these LLMs~\cite{tian2023chatgpt,xia2023automated,li2023exploring,aiyappa2023can}. For example, Tian~\etal~\cite{tian2023chatgpt} discovered that ChatGPT's correctness on {\em 2022 LeetCode questions} was significantly lower than in previous years, where the questions were available online when ChatGPT data was being collected (i.e., before September 2021). This finding suggests that, to ensure more accurate assessment of LLMs, careful benchmark selection is essential to reduce data leakage risks.
\item[\ding{173}] {High Computational Overhead}: 
 Current interactive repair methods can be computationally intensive. For instance, Xia~\etal~\cite{xia2023conversation}'s ChatRepair requires an average of 10 independent dialog sessions to repair an incorrect code [15]. This high computational cost can limit the practicality of these approaches.
\item[\ding{174}] {Limited Evaluation on Large-Scale Datasets}: To the best of our knowledge, existing interactive repair methods have not been evaluated on large-scale datasets. This includes recent research by Xia~\etal~\cite{xia2023conversation}, Azaria~\etal~\cite{azaria2023chatgpt}, and Sobania~\etal~\cite{sobania2023analysis}. Assessing the performance of these state-of-the-art methods on larger datasets would provide a more comprehensive understanding of their capabilities and limitations.

\item[\ding{175}] {Need for Augmented Information}: Existing LLM-based interactive program repair methods primarily rely on problem descriptions and failing test cases to generate patches~\cite{xia2023conversation,xia2023conversational}. However, this reliance might not give LLMs enough information to understand the programs and identify the necessary fixes. Prior research~\cite{bohme2020human,azaria2023chatgpt,winter2022let} indicated that incorporating human interaction into the repair process can significantly improve the quality and accuracy of generated patches. Furthermore, leveraging diverse augmented information, such as solution descriptions, can deepen LLMs' understanding of programs~\cite{bohme2020human,goues2019automated,monperrus2018automatic,yang2024multiobjective}. Consequently, it is necessary to investigate approaches to integrating augmented information with human expertise, specifically \textbf{interaction with tutors or developers}, into the repair process, further enhancing LLMs' repair capabilities.
\end{itemize}

\noindent {\bf This paper.} 
In this study, we address the challenge of data leakage and evaluate the practical benefits of different forms of augmented information within conversation-based program repair methods. To do this, we introduce an extensive benchmark dataset, which we refer to as ``{\em uncrawled}'' to emphasize that it has not been incorporated into the training data of any pre-trained LLM. This benchmark originates from a company specializing in training novice developers to become experienced professionals through data structure and algorithm courses.

Our dataset, \dataset, comprises 1,239 incorrect C++ code samples contributed by 427 students, covering 35 distinct programming challenges distributed across 12 difficulty levels. It also includes tutor guidance provided by human tutors as well as corresponding corrected code. The 35 challenges were designed to cultivate various cognitive skills, including abstract reasoning, procedural thinking, and conceptual understanding. Sourced from the real-world experience of repair processes, we expect this benchmark to serve as a valuable and reliable asset\footnote{\dataset will be made available due to our commitment to open science. However, to prevent web crawlers from retrieving it (with the risk of it being leaked into training data of LLMs), we restrict its access via an API requiring authorization tokens.} for evaluating a range of coding-related tasks, such as program synthesis, fault localization, and program repair.
Using \dataset, we explore the realistic repair capabilities of LLMs and investigate LLMs as conversational repair tools for tutors in programming education scenarios.
First, we investigate the practical program repair capabilities of 12 prominent LLMs. Our analysis reveals that GPT-4 and GPT-3.5 consistently outperform other LLMs in program repair tasks. Second, we explore the influence of different types of augmented information in prompts on LLM-based program repair performance. Our findings demonstrate that providing LLMs with tutor guidance significantly enhances their performance, and this can be further improved by incorporating solution descriptions and failing test cases. Finally, to leverage the conversational abilities of LLMs and capitalize on the aforementioned three types of augmented information, we introduce a novel semi-automatic LLM-based {\bf C}onversational program {\bf \textsc{RE}}pair {\bf F}ramework~(\tool) for tutors, prompting LLMs with tutor guidance, solution description, and failing test cases, interactively. 
Leveraging LLM conversational capabilities, \tool achieves an AVG-5 score of 76.6\% when utilizing GPT-4, showcasing remarkable program repair capabilities with only  7.5\% of input tokens~(\ie expense cost) per request compared to ChatRepair. In practical application within a company, \tool acts as a semi-automatic repair tool that aids tutors by utilizing LLMs to repair incorrect codes based on preliminary feedback, thus reducing debugging times by 71.2\% and decreasing costs by 69.9\%. \tool enhances the tutoring effectiveness and improves the learning experience by providing more timely and accurate debugging assistance.

\noindent 
\textbf{Contributions}. The main contributions of our work are as follows:

\begin{itemize}[noitemsep,topsep=0pt,leftmargin=*]

\item We introduce a large-scale uncrawled benchmark \dataset for realistic evaluation of LLM-based repair approaches. \dataset includes 1,239 defective C++ programs, to which human tutor guidance, programming problem description, solution description, test cases, and ground truth corrected codes are attached. 

\item We assess the realistic repair capabilities of state of the art 8 open-source~\cite{Tunstall2023starchat-alpha,nijkamp2022codegen,fried2022incoder,replit-code,zheng2023vicuna,roziere2023codellama} and 4 closed-source~\cite{openai2023gpt4,ouyang2022training,anthropic2023claude,pichai2023bard} LLMs on the \dataset.

\item We evaluate the enhancements of different augmented information on the repair capabilities of LLMs, measuring repair correctness~(AVG-5) and patch precision~(RPSR).

\item We introduce a conversational semi-automatic repair framework, termed \tool, to leverage the conversational capabilities of any LLMs and different augmented information for repair tasks. The effectiveness of \tool is validated using \dataset.

\item We deploy \tool as an \textbf{assisting tool for programming tutors} in a company, achieving a 71.2\% response time reduction and a 69.9\% cost decrease for students' debugging requests.

\end{itemize}

The structure of this paper is organized as follows: Section \ref{background} offers a review of background and related works. In Section \ref{studydesign}, we design the methodology employed in this study. Section \ref{experiments} provides a detailed account of the experimental findings, while Section \ref{discussion} investigates the outlier data and demonstrates the industrial application. Potential threats to the validity are qualified in Section \ref{threats}. Section \ref{conclusion} summarizes this paper's key points and findings.

\section{Background \& Related Work}
\label{background}

\subsection{Large Pre-Trained Language Model}

\noindent\textbf{Encoder-Decoder Models}

Encoder-decoder architectures, such as BART~\cite{lewis2020bart} and T5~\cite{raffel2020t5}, employ an encoder to transform an input sequence into a continuous representation, capturing essential information. A decoder then generates the output sequence based on this representation. \textbf{CodeT5p}~\cite{wang2023codet5+}, an advancement of CodeT5~\cite{wang2021codet5}, is developed by Salesforce and is pre-trained on a wide array of tasks, incorporating both unimodal code data and bimodal code-text data.




\noindent\textbf{Decoder-Only Models}

Decoder-only architectures, such as GPT-3~\cite{brown2020gpt3} and CodeGen~\cite{nijkamp2022codegen}, utilize the transformer's decoder in an autoregressive manner to sequentially generate tokens based on preceding tokens. This approach is the most prevalent transformer architecture at present.

The \textbf{G}enerative \textbf{P}re-trained \textbf{T}~(GPT) model by OpenAI features a unidirectional, causal attention mechanism. This design ensures that each sequence position attends only to preceding positions, thereby maintaining the input sequence order. During pre-training, the model maximizes the likelihood of word predicting based on its preceding context, as denoted by the following formula:

\vspace{-0.2cm}
\[
Likelihood = \sum_{t} \log P(w_t | w_{1:t-1}; \theta)
\]
\vspace{-0.2cm}

Both Codex~\cite{chen2021codex} and InstructGPT~(\textbf{GPT-3.5})~\cite{ouyang2022training} are derivatives of the GPT-3 model. Additionally, ChatGPT~\cite{chatgpt} served as a dialog-optimized adaptation of GPT, developed through Reinforcement Learning from Human Feedback~(RLHF) and drawing from the frameworks of both Codex and InstructGPT. The successor to GPT-3.5, known as \textbf{GPT-4}~\cite{openai2023gpt4}, boasts enhanced performance and an extended context window. GPT-3.5 and GPT-4 are proficient in code generation and debugging, guided by natural language conversational contexts~\cite{tian2023chatgpt}. \textbf{Claude-instant}~\cite{anthropic2023claude}, based on Anthropic's Constitutional AI~\cite{bai2022constitutional}, is a next-generation AI assisting trained on an extensive corpus of text and code thus capable of code-related tasks~\cite{claudeAI2023}. \textbf{Bard}~\cite{pichai2023bard}, Google's experimental conversational AI, based on LaMDA~\cite{thoppilan2022lamda}. It is pre-trained on a 1.56 TB dataset, comprising public dialog data and web documents, which includes 12.5\% of code-related documents. \textbf{StarChat}~\cite{Tunstall2023starchat-alpha}, released by HuggingFaceH4, is a derivative of BigCode's StarCoder. It is pre-trained as a Transformer decoder-only model named StarCoderBase and is subsequently fine-tuned for Python coding tasks. Salesforce's \textbf{CodeGen}~\cite{nijkamp2022codegen} employs a multi-step paradigm for program synthesis that outperforms single-turn methods. \textbf{Incoder}~\cite{fried2022incoder}, developed by Meta, uses a causal-masked objective for training and specializes in code insertion and code generation. \textbf{Replit-code}~\cite{replit-code}, a 2.7B Causal Language Model, focuses on code completion and employs Flash Attention and AliBi positional embeddings for efficient training and inference. LMSYS's \textbf{Vicuna}~\cite{zheng2023vicuna} is a chat assistant fine-tuned from Llama~\cite{touvron2023llama} utilizing shared conversational data. Lastly, \textbf{CodeLlama}~\cite{roziere2023codellama} represents a leading family of LLMs tailored for coding tasks, with multiple versions including foundational models, Python specializations, and instruction-following models. Given the numerous LLMs that can repair code, we list 12 prominent LLMs in Section \ref{subsec:models} for subsequent experimental purposes.

\subsection{Interactive Program Repair}

In the context of program repair, ``patch'' denotes the difference between the original and corrected code. Test cases are employed to verify the integrity of patches produced by repair tools. The ``patch size'' represents the magnitude of modifications in the altered code. Various metrics can quantify this size, including line differences~\cite{wang2019attention}, alterations in the Abstract Syntax Tree~(AST)~\cite{DBLP:refactoring,gulwani2018automated}, and other measures, each offering a unique perspective on the complexity of the modifications.

Given the LLM's inherent unpredictability and opacity in coding tasks, engaging human interactions in a dialogic collaboration becomes imperative for obtaining dependable results~\cite{azaria2023chatgpt, ge2023llm}. Gao~\etal~\cite{gao2020interactive} introduced an interactive repair methodology, enabling developers to review and select patches generated by automated techniques. The approach translated the patch into interrogative frameworks, simplifying the process for developers who can then make informed selections without delving into the intricate semantics of the patch. Sobania~\etal~\cite{sobania2023analysis} explored the utility of dialog-based interactions in ChatGPT. Empirical studies conducted on the QuixBugs~\cite{DBLP:Quixbugs} demonstrated the efficacy of dialogic hints in enhancing the program repair capabilities. Xia~\etal~\cite{xia2023conversation} proposed ChatRepair, a paradigm that prompts ChatGPT with generated incorrect patches and failing test cases. In their approach, each repair case consisted of approximately 10 distinct conversation sessions, with each conversation limited to a maximum of three turns. Experiments on 10 LLMs showed improvements, validating the effectiveness of providing validation feedback in a conversational manner for program repair. In our work, we investigate the potential of the interactive capabilities of LLMs to enhance LLM-based program repair.

\subsection{Intelligent Tutoring System for Programming}

In education scenarios, \textbf{I}ntelligent \textbf{T}utoring \textbf{S}ystem (ITS) for programming shows promising results in helping novice students learn programming~\cite{DBLP:refactoring,gulwani2018automated,fan2023automated,ahmed2022verifix,zhang2022repairing}. Gulwani~\etal proposed Clara~\cite{gulwani2018automated}, a fully automated program repair algorithm for introductory C/Python programming assignments, using the existing correct student solutions to repair the incorrect attempts through clustering. Yang~\etal proposed Refactoring~\cite{DBLP:refactoring}, a fully automated approach for generating student Python program repairs in real-time, using re-factoring rules to generate a correct solution with the same control flow as the incorrect program. ITSP~\cite{yi2017feasibility} generates partial repairs that serve as hints to guide students toward the reference solution, achieving a 57\% repair rate on a dataset of 661 incorrect C programs collected from an introductory programming course. Verifix~\cite{ahmed2022verifix} is an automated program repair technique for introductory C programming assignments that generates repairs by aligning a student's incorrect program with a reference solution using control flow. MMAPR~\cite{zhang2022repairing} was proposed to use Codex~\cite{chen2021codex} with few-shot prompting to build an APR system for introductory Python programming assignments. The evaluation results on 286 real student programs show that MMAPR can fix more programs and produce smaller patches than Refactoring. Yasunaga~\etal proposed DrReapir~\cite{yasunaga2020graph}, a graph-based program repair technique that learns to repair bugs in programming assignments by leveraging diagnostic feedback from a compiler, employs a self-supervised learning paradigm that generates extra training data by corrupting unlabeled programs and obtains diagnostic feedback. Fan~\etal's study revealed that given bug location information provided by a statistical fault localization approach, Codex with edit mode is similar to or better than existing Java state-of-the-art repair tools in fixing incorrect codes~\cite{fan2023automated}.

Due to the unavailability of the closed-source Codex used by Fan~\etal's approach as well as MMAPR, and the lack of C++ language support in Clara, ITSP, Verifix, and Refactoring, this paper focuses on evaluating the repair capabilities of DrRepair.
\section{Study Design}
\label{studydesign}

\subsection{Research Questions}
\begin{itemize}[noitemsep,topsep=0pt,leftmargin=*]
\item \textbf{RQ-1}: \textbf{\textit{How effective are state-of-the-art LLMs in repairing incorrect code?}} We evaluate the realistic performance of 12 prominent LLMs in repairing code based on \dataset (an uncrawled dataset that enables fair and unbiased evaluation of LLMs for code repair). We assess both the correctness as well as the precision of the generated patches. A baseline for evaluation is established using input data consisting of incorrect code and associated programming problem description. Furthermore, this study delves into the influence of code length and difficulty of programming tasks on the repair capabilities of LLMs.

\item \textbf{RQ-2}: \textbf{\textit{Can augmented information assist in strengthening the repair capabilities of LLMs?}} This study employs three types of augmented information to enhance the repair capabilities of LLMs: \textit{solution description} as general hints, \textit{tutor guidance} as specific guidance, and \textit{failing test cases} to provide automated testing feedback. The study provides a thorough analysis of the impact of various combinations of augmented information on the repair capabilities of LLMs.

\item \textbf{RQ-3}: \textbf{\textit{To what extent can conversation-based repair further exploit the repair capabilities of LLMs?}} Utilizing the conversational strengths of LLMs and incorporating human interactions, this study introduces \tool, a semi-automatic conversational repair framework. A large-scale automated experiment is conducted using the \dataset to evaluate its effectiveness. \tool incorporates three types of augmented information to unlock the potential repair capabilities of LLMs through multiple rounds of dialogues. The study validates the efficacy of conversation-based repair methods by comparing the performance disparities when historical dialogue entries are included or excluded in each conversation turn.

\end{itemize}

\subsection{Benchmark Selection Criteria}
In this paper, C++ is selected as the repair benchmark's programming language. Given its widespread application in high performance and interfaces, including game development, large-scale data platforms, and operating systems~\cite{jiang2022generation,stroustrup2013c++}. Furthermore, C++ is a multi-paradigms programming language, posing additional complexities for program repair tools~\cite{stroustrup2013c++}.

\subsection{Dataset}
\label{subsec:dataset}

\begin{figure*}[ht]
\centering
\begin{minipage}[t]{0.9\textwidth}
\begin{subfigure}[t]{0.45\textwidth}
\centering
\begin{lstlisting}[language=mymarkdown,frame=tb,mathescape=true]
  # Artificial Intelligence

  In order to make users believe they are talking to an $AI$ instead of a real person, it is not an easy task to take on this position. Mr. Garlic is one of the best in this role, and his performance is outstanding.
  The company's top management praised Mr. Garlic's abilities and organized a learning activity for him to share his work experience. However, when everyone came to Mr. Garlic's workstation, they found no one in the chair, and only saw a program running on the computer:
  Accept user input, then:
    1. Change all capital English letters in the original text to lowercase, except for `I`;
    2. Replace all standalone `can you` and `could you` in the original text with `I can` and `I could`, respectively;
    3. Replace all standalone `I` and `me` in the original text with `you`;
    4. Convert `?` to `!`;
  Finally, output the result as the reply to the user. Can you understand this program?
  ### Input Format
  The first line contains an integer $n$, indicating that there will be $n$ operations. The next $n$ lines contain a string representing user input ($len \le 100$, no more than 100 lines).
  ### Output Format
  For each user input, output a corresponding line of string, representing the AI's output content.
  ### Time Limit
  1000ms
  ### Memory Limit
  65536KB
\end{lstlisting}
\caption{Programming problem description}
\label{fig:desc_eg}
\end{subfigure}
\hfill
\begin{minipage}[t]{0.5\textwidth}
\begin{subfigure}[t]{\textwidth}
\centering
\begin{lstlisting}[language=mymarkdown,frame=tb]
Since line 30 has judged f[i+1], it should let i++ and skip f[i+1].
Also, note that there are multiple sets of data, and you need to output for each input. Remember to clear the f array.
\end{lstlisting}
\caption{Tutor guidance}
\label{fig:def_reply_eg}
\end{subfigure}
\begin{subfigure}{\textwidth}
\centering
\begin{lstlisting}[language=mymarkdown,frame=tb,mathescape=true]
Use `getline()` to read each line of the string, and use the `isalnum` function to determine whether it is a number or a letter. If it is a number or a letter, a non-$0$ value is returned. If it is, check if it is an `I`, as we need to convert it to lowercase. If it is not a number or a character, insert a space (the reference code for handling spaces uses `stringstream`, but other methods can also be used). After processing the spaces, split the words in the sentence for judgment. Pay attention to the spaces before and after the words and the ultimate string operation. Pay attention to the details, and remember to output the original sentence first!
\end{lstlisting}
\caption{Solution description}
\label{fig:doc_eg}
\end{subfigure}
\begin{subfigure}{\textwidth}
\centering
\begin{lstlisting}[language=mymarkdown,frame=tb]
[INPUT]      5
                     Can YoU aNswer me ?
                     ......
[OUTPUT]   I can answer you !
                     ......
\end{lstlisting}
\caption{Failing test cases}
\label{fig:def_code_testcase}
\end{subfigure}
\end{minipage}
\end{minipage}
\caption{Examples of programming problem description and three types of augmented information.}
\label{fig:eg}
\end{figure*}

LLMs like ChatGPT often incorporate code from various sources, including GitHub, into their training datasets~\cite{openai2023gpt4,brown2020gpt3,roziere2023codellama}. Most C++ repair benchmarks, such as IntroClass~\cite{DBLP:introclass} and ITSP~\cite{yi2017feasibility}, are hosted on GitHub. Additionally, benchmarks like Bugs-C++~\cite{an2023bugsc++} and ManyBugs~\cite{DBLP:introclass} compile various historically popular open-source projects from GitHub. Qingyuan~\etal highlighted that frequently incorporating GitHub-hosted program repair benchmarks into conventional training datasets leads to data leakage and subsequent inaccuracies in performance evaluation~\cite{Qingyuan2024empirical}. This suggests a widespread potential for data leakage across existing C++ repair benchmarks. To mitigate the risk of data leakage during our experimental analysis, we introduce a large-scale uncrawled C++ repair benchmark, denoted as \dataset. \dataset originates from proprietary, internal data collated by a company, specifically from data structure and algorithm courses aimed at training novice developers to experts. \dataset is publicly available through API to mitigate the risk of unauthorized crawling into the training corpus of LLMs. Under the usage license, API users are strictly limited to uploading \dataset to the public network. \dataset comprises 1,239 incorrect codes written by 427 students and covers 35 programming problems, adapted from real-world development context to single-task problems assessed using standardized input-output test cases. Each incorrect code is accompanied by problem description~(Figure~\ref{fig:desc_eg}), tutor guidance~(Figure~\ref{fig:def_reply_eg}), solution description~(Figure~\ref{fig:doc_eg}), and failing test cases~(Figure~\ref{fig:def_code_testcase}). The public benchmark \dataset will be \textbf{continuously} updated with new problems and buggy codes in the future, and will serve as a fundamental resource for future code-related research.

\vspace{-0.2cm}
\begin{table}[H]
\caption{12 Tiers of \dataset.}
\label{tab:difficulty}
\centering
\resizebox{.8\columnwidth}{!}{%
\begin{tabular}{llll}
\hline\noalign{\smallskip}
Tiers Range & Summary & Key Objectives \\
\noalign{\smallskip}\hline\noalign{\smallskip}
T1, T2, T3 & Introduction to C++ Grammar & Abstract Thinking, \\
 & & Program Thinking \\
\noalign{\smallskip}\hline
T4, T5, T6 & Introduction to Data Structures & Object Thinking, \\
 & and Basic Algorithms & Relationship Modeling \\
\noalign{\smallskip}\hline
T7, T8, T9 & Common Data Structures & Algorithm Modeling, \\
 & and Algorithms & Abstract Stateful Thinking \\
\noalign{\smallskip}\hline
T10, T11, T12 & Advanced Data Structures & Time-space Trade-off, \\
 & and Algorithms & Mathematical Deduction \\
\noalign{\smallskip}\hline
\end{tabular}
}
\end{table}
\vspace{-0.2cm}

\begin{table}[h]
\caption{Statictics of \dataset.}
\label{tab:data_set}
\centering
\resizebox{.9\columnwidth}{!}{%
\begin{threeparttable}
\begin{tabular}{cccccc}
\hline\noalign{\smallskip}
Title & Category & Tier & Codes\tnote{1} & Lines\tnote{2} & Hunks\tnote{3}  \\
\noalign{\smallskip}\hline\noalign{\smallskip}
Absolute Value Sorting  &  Branch  & T1 & 85 & 29.5 & 2.9\\
Number of Days in a Month  &  Branch  & T1 & 64 & 20.5 & 2.6\\
Binary Tree Sorting  &  Binary Tree  & T10 & 36 & 45.6 & 4.9\\
Magical Key  &  Char  & T2 & 81 & 23.8 & 3.0\\
Receipt  &  Loop  & T2 & 15 & 32.6 & 2.0\\
Chasing the Enemy  &  Loop  & T3 & 88 & 15.1 & 2.1\\
Compression Technology  &  Array  & T3 & 22 & 33.3 & 2.9\\
Find the Longest Word  &  Loop  & T3 & 67 & 24.9 & 3.7\\
Advanced Integer Sorting  &  Sort  & T4 & 66 & 32.8 & 3.1\\
Gem Collector II  &  Sort  & T4 & 26 & 46.3 & 4.0\\
Maximum Submatrix  &  Enumeration  & T4 & 33 & 36.7 & 3.2\\
Artificial Intelligence  &  Ad Hoc  & T5 & 63 & 41.5 & 4.4\\
Onion Girl's Flying Chess  &  Recursion  & T5 & 20 & 29.4 & 4.2\\
Annoying Queue Jumping  &  Queue  & T6 & 60 & 41.6 & 3.3\\
Complex Stacks  &  Stack  & T6 & 22 & 74.6 & 5.9\\
Talent Show  &  Vector  & T6 & 54 & 35.0 & 3.9\\
Four Squares Theorem  &  Enumeration  & T7 & 30 & 25.5 & 3.5\\
High Precision Factorial  &  Big Integer  & T7 & 11 & 46.8 & 4.4\\
p Nodes  &  Tree Structure  & T7 & 48 & 40.3 & 4.4\\
Perfect Match  &  Prefix Sum  & T7 & 15 & 31.5 & 3.3\\
Program Design T2 &  Binary Search  & T7 & 8 & 31.9 & 3.4\\
Program Design T3 &  Big Integer  & T7 & 23 & 27.0 & 2.8\\
2n Queens Problem &  DFS  & T8 & 37 & 52.3 & 4.9\\
Elevator  &  Maths  & T8 & 26 & 31.3 & 5.1\\
Going Outdoors  &  DFS  & T8 & 19 & 47.3 & 2.8\\
Information Parser  &  Maths  & T8 & 34 & 65.4 & 6.1\\
Super Bookshelf 2 &  DFS  & T8 & 14 & 27.9 & 2.7\\
Escape  &  DP  & T9 & 24 & 74.2 & 5.8\\
I Want to Stay Healthy Today  &  DP  & T9 & 33 & 38.4 & 4.0\\
Noble Shops  &  DP  & T9 & 8 & 33.5 & 5.1\\
Furious Stones  &  DP  & T10 & 30 & 30.0 & 3.4\\
Maze  &  BFS  & T10 & 20 & 76.2 & 9.3\\
Foodie Mr. Garlic  &  DP  & T11 & 23 & 47.6 & 5.6\\
Highways  &  Graph Theory  & T12 & 13 & 104.8 & 13.3\\
Mr. Garlic's Treasure Hunt  &  Graph Theory  & T12 & 21 & 145.6 & 10.3\\
\noalign{\smallskip}\hline
\end{tabular}
\begin{tablenotes}
      \small
      \item [1] ``Codes'' refers to the number of incorrect codes.
      \item [2] ``Lines'' refers to the average lines of incorrect codes.
      \item [3] ``Hunks'' refers to the average changed hunks between incorrect codes and corrected codes.
    \end{tablenotes}
\end{threeparttable}
}
\end{table}

All the programming problems in \dataset are derived from real-world software development contexts and have been curated by 20 senior software experts. As illustrated in Figure \ref{fig:desc_eg}, each problem description includes vital elements such as the title, task objective, input-output formats, and computational constraints. The constraints describe the accepted data formats and ranges of input values, while the limitations impose bounds on execution time and memory usage. The problems are comprehensive, covering the common data structures and algorithms, allowing for a thorough evaluation of LLMs' code capabilities across diverse software development scenarios. For each programming problem, \dataset contains 5-10 sets of paired input-output test cases, as detailed in Figure \ref{fig:def_code_testcase}. Since 2017, these test cases have been continually refined to ensure their quality. Tutor guidance is provided as targeted hints after a code review without revealing the corrected code, as shown in Figure \ref{fig:def_reply_eg}. Solution descriptions offer a high-level approach rather than specific code implementations, depicted in Figure \ref{fig:doc_eg}.

Contrastingly, \dataset is uniquely structured into 12 difficulty tiers, each labeled by software developer experts. These tiers are further subdivided into four stages. The objectives for each stage are outlined in Table \ref{tab:difficulty} and correspond to critical cognitive stages in the professional development of novice developers. Each tier has at least two programming problems and a minimum of 34 incorrect codes. A hierarchical arrangement of 20 knowledge tags, organized by tier, is provided in Table \ref{tab:data_set}. As one ascends the difficulty tier, there is a corresponding increase in algorithmic complexity and the intricacy of code implementation. On average, each programming problem in \dataset includes 7.27 domain expert-designed test cases that have been consistently evaluated as sufficiently valid for assessing code correctness from 2017 to the present. To avoid flaky tests for the repaired codes, we offer a public testing API for each programming problem within \dataset, facilitating the direct acquisition of results and eliminating environmental instabilities.

We have preprocessed incorrect codes by removing duplicates disregarding whitespace differences, to ensure the uniqueness of bugs within \dataset. Compared to widely-used Java repair benchmark Defects4J~\cite{DBLP:defects4j}, \dataset exhibits significant assessment diversity, not only in the availability of tutor guidance information but also in terms of bug complexity. Only 18.6\% of bugs in \dataset involve one modified hunk versus 64.2\% in Defects4J. \dataset spans a wider array of complexities, from one to more than six modified hunks. Additionally, in \dataset, 47.8\% of incorrect codes contain multiple functions, and 38.4\% of incorrect codes exhibit defects across multiple functions, while all the buggy codes in Defects4J are single-function modifications. \dataset provides a foundation for future investigations into the repair capabilities of tools across various complexities.

\subsection{Evaluation Metrics}
\label{subsec:metrics}

Given LLM's inherent \textit{\textbf{randomness}}, multiple requests with an identical prompt may yield different outputs. Multiple requests are made to LLMs using the same prompt to mitigate this randomness and facilitate robust analytical outcomes. Prior research has shown that typically, five instances are generated using LLMs for each incorrect code.

\noindent\textbf{1. Repair Correctness.} This study evaluates the correctness of LLM-based repairs through two key quantitative metrics: TOP-5 and AVG-5. \textbf{TOP-5} quantifies the likelihood that LLMs will produce at least one accurate repair out of five attempts. This measurement is designed to align with the highest number of code revisions most developers are willing to review~\cite{kochhar2016practitioners}. Additionally, \textbf{AVG-5} is a more stable metric than the somewhat unpredictable \textbf{TOP-1} that LLMs produce; it reflects the mean number of accurate repairs across five trials.

\noindent\textbf{2. Patch Precision.} In \textbf{programming education scenarios}, minimizing code changes is the key to preserving intent for students, aiding their understanding of how to fix bugs~\cite{shirafuji2023program,gulwani2018automated}. Existing program repair tools often generate larger size patches that deviate from the original codes~\cite{wang2019different}, while optimal patches should be minimalistic, addressing code defects without introducing unnecessary modifications~\cite{dong2020priority}. Therefore, evaluating the size of patches generated by LLMs becomes vital in assessing the precision of LLM-based repair methods. Gulwani~\etal~\cite{gulwani2018automated} introduced the Relative Patch Size~(RPS) as a metric to quantify patch size. RPS is formulated as follows:

\[
RPS(AST_i, AST_r) = \frac{TED(AST_i, AST_r)}{Size(AST_i)}
\]

Here, $AST_{i}$ and $AST_{r}$ denote the Abstract Syntax Tree~(AST) of the incorrect code and LLM-generated repaired code, respectively, while $TED(AST_i$, $AST_r)$ is the Tree-Edit-Distance~(TED) between them. $Size(AST_i)$ represents the AST node count of the incorrect code. According to the formula, RPS scores can exceed 1.0 and approaching infinity for null programs, and they are inherently affected by the distribution of successful repairs. This study introduces a new metric, the Relative Patch Size Ratio~(\textbf{RPSR}), to address this limitation and utilize ground truth corrected codes. RPSR is defined as:

\[
RPSR(AST_i, AST_c, AST_r) = \frac{RPS(AST_i, AST_r)}{RPS(AST_i, AST_c)} = \frac{TED(AST_i, AST_r)}{TED(AST_i, AST_c)}
\]

Where $AST_c$ means the abstract syntax tree of ground truth corrected code. For each LLM-fixed code, a lower RPSR means a more precise patch, while RPSR values lower than 1.0 indicate smaller patch sizes produced by LLMs compared to ground truth patches.

\subsection{Models}
\label{subsec:models}

This study evaluates twelve state-of-the-art LLMs, divided into two categories: eight open-source LLMs and four closed-source LLMs. The LLMs are selected based on the number of downloads from official repositories hosted by HuggingFace. The open-source LLMs are shown in Table \ref{tab:open_llm}, featuring parameter sizes ranging between 6b and 16b. Within this table, the column labeled \textit{\#Param} specifies the size of the model parameters, and the \textit{Downloads} column provides the aggregate number of downloads for each LLM up to September 2023 across all sub-categories. The \textit{HumanEval TOP-1} column presents the TOP-1 performance metric for each LLM, as measured by the HumanEval~\cite{chen2021codex} benchmark.

\begin{table}[h]
\caption{8 selected open-source LLMs~(Sept. 2023).}
\label{tab:open_llm}
\centering
\resizebox{.8\columnwidth}{!}{
\begin{tabular}{ccccc}
\hline\noalign{\smallskip}
Model & Training Dataset & \#Param  & Downloads & HumanEval TOP-1  \\
\noalign{\smallskip}\hline\noalign{\smallskip}
CodeLlama-instruct-13B & N.R. & 13B & 522.6k & 42.7~\cite{roziere2023codellama} \\
Vicuna-13B & BigQuery & 13B  & 440.2k & 15.5~\cite{liu2023your}\\
CodeGen-6B / 16B & BigQuery & 6B / 16B  & 91.1k & 27.7 / 32.2~\cite{liu2023your}\\
StarChat-alpha & Stack Dedup v1.2 & 16B & 82.6k & 30.0~\cite{babe2023studenteval}\\
CodeT5p-16B & CodeSearchNet & 16B  & 21.7k & 30.9~\cite{luo2023wizardcoder}\\
Incoder-6B & N.R.  & 6.7B  & 12.4k & 15.6~\cite{liu2023your}\\
Replit-code-v1 & Stack Dedup v1.2  & 2.7B & 2.3k & 21.9~\cite{replit-code}\\
\noalign{\smallskip}\hline
\end{tabular}
}
\end{table}

Table \ref{tab:close_llm} enumerates the closed-source LLMs, with the \textit{Institution} column describing each model to its respective affiliated organization. The remaining columns in this table shared the same attributes as those outlined in Table \ref{tab:open_llm}.

\begin{table}[h]
\caption{4 selected closed-source LLMs.}
\centering
\label{tab:close_llm}
\resizebox{.8\columnwidth}{!}{%
\begin{tabular}{ccccc}
\hline\noalign{\smallskip}
Model & Institution & \#Param  & Usage Type & HumanEval TOP-1 \\
\noalign{\smallskip}\hline\noalign{\smallskip}
GPT-4-0613-8k & OpenAI & 1800B & API / Web & 67.0~\cite{openai2023gpt4} \\
GPT-3.5-Turbo-0613-4k & OpenAI & 175B & API / Web & 48.1~\cite{openai2023gpt4} \\
Claude-instant-v1 & Anthropic & 52B & Web & 47.6~\cite{luo2023wizardcoder} \\
Bard & Google & 137B & Web & 44.5~\cite{luo2023wizardcoder} \\
\noalign{\smallskip}\hline
\end{tabular}
}
\end{table}

In the subsequent experiments, specific abbreviations are used for ease of reference: GPT-4 to GPT-4-0613-8k, GPT-3.5 to GPT-3.5-Turbo-0613-4k, Claude to Claude-instant-v1, StarChat to StarChat-alpha, CodeLlama to CodeLlama-instruct-13B. The sampling temperature for these selected LLMs is set to a consistent value of 1.0, which aligns with the default settings employed across most LLMs.

\subsection{Prompts and Augmented Information}
\label{subsec:prompts}

\begin{figure}[H]
\centering
\begin{minipage}{.9\columnwidth}
\centering
\captionsetup{justification=centering}
\begin{lstlisting}[language=mymarkdown,frame=tb,basicstyle=\footnotesize]
  This is a programming problem description:
  {{description}}
  This is an incorrect code to the problem:
  {{incorrect code}}
  You are a software engineer. Can you repair the incorrect code?
\end{lstlisting}
\end{minipage}
\caption{The prompt format of the baseline.}
\label{fig:prompt_baseline}
\end{figure}

The performance of LLMs in repair tasks is significantly influenced by the prompt's design~\cite{cao2023study,tian2023chatgpt,kemker2018measuring}. The optimal prompt format, depicted in Figure \ref{fig:prompt_baseline}, mitigates the adverse impact of lengthy prompts. This format includes a problem description, an incorrect code encased in triple backticks, and a task prompt.

\begin{figure}[h]
\centering
\includegraphics[width=\columnwidth]{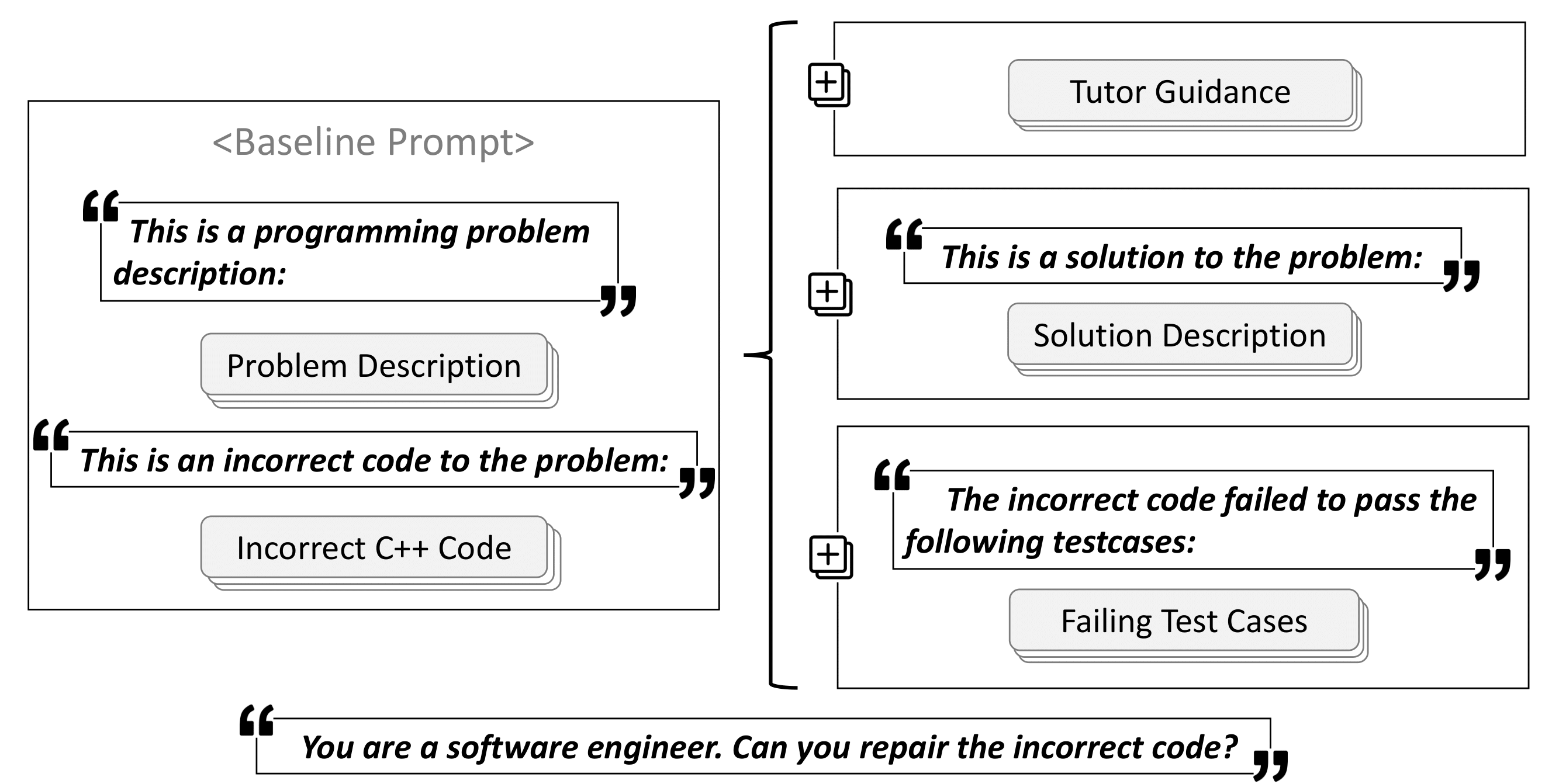}
\caption{The baseline's prompt structure entails three augmented information types.}
\label{fig:prompts}
\end{figure}

To assess the impact of augmented information on LLM's repair capabilities, comprising tutor guidance, solution description, and failing test cases, a uniform prompt structure was used in the following experiments to avoid sensitive prompts for LLMs. This framework, displayed in Figure \ref{fig:prompts}, initiates with a problem description and incorrect code consistent with the baseline prompt. It then incorporates one or more types of augmented information. The solution description commences with the phrase \textit{"This is a solution to the problem:"}, followed by the complete content of the solution description. In the context of tutor guidance, content is explicitly provided. When presenting failing test cases, the framework employs the introductory phrase \textit{"This incorrect code failed to pass the following test cases:"} succeeded by the corresponding failed input-output pairs. The concluding task description retains alignment with the baseline, thus facilitating a rigorous comparative analysis of the impact of augmented information on the repair performance of LLMs.
\section{Experiment \& Result}
\label{experiments}

\subsection{Realistic Repair Performance of LLMs}
\noindent\textbf{[Experiment Goal]:}
We aim to investigate the realistic repair performance of 12 state-of-the-art LLMs and an existing ITS technique, utilizing the \dataset as our benchmark.

\noindent\textbf{[Experiment Design]:}
This experiment provides LLMs with prompts in the baseline format, as described in Section \ref{subsec:prompts}. The repair capabilities of LLMs are assessed in terms of three metrics: TOP-5, AVG-5, and RPSR, as detailed in Section \ref{subsec:metrics}. The AVG-5 results of 5 best-performing LLMs across 12 tiers are calculated to compare the repair performance of LLMs across various tiers and analyze their trends. Additionally, we analyze the impact of code length on the repair capabilities of LLMs. We calculate the boxplot distribution of the length of incorrect codes for correct and incorrect predictions generated by LLMs.

\begin{table}[H]
\caption{TOP-5, AVG-5, RPSR results of 12 selected LLMs on \dataset.}
\label{tab:2}
\resizebox{.8\columnwidth}{!}{
\begin{threeparttable}
\centering
\begin{tabular}{ccccc}
\hline\noalign{\smallskip}
Model & TOP-5 & AVG-5 & H-TOP-1\tnote{*}  & RPSR   \\
\noalign{\smallskip}\hline\noalign{\smallskip}
GPT-4 & 66.5\% & 52.0\% & 67.0\% & 3.748 \\
GPT-3.5 & 56.8\% & 41.5\% & 48.1\% & 5.072 \\
Claude & 34.1\% & 20.3\% & 47.6\%  & 4.083  \\
Bard & 27.9\% & 16.8\% & 44.5\% & 4.728  \\
\noalign{\smallskip}\hline\noalign{\smallskip}
CodeLlama & 15.9\% & 6.8\% & 42.7\% & 3.635 \\
StarChat & 11.7\% & 5.7\% & 30.0\% & 8.028 \\
Vicuna-13B & 1.4\% & 0.5\% & 15.5\% & 6.140\\
CodeGen-multi-16B & 0.8\% & 0.3\% & 32.2\% & 1.745\\
CodeGen-multi-6B &  0.0\% & 0.0\% & 27.7\% & / \\
CodeT5p & 0.3\% & 0.1\% & 30.9\% & 1.013 \\
Incoder & 0.2\% & 0.1\% & 15.6\% &  0.860  \\
replit-code-v1 & 0.1\% & 0.1\% & 21.9\% & 0.959 \\
\noalign{\smallskip}\hline\noalign{\smallskip}
DrRepair & 0.2\% & 0.2\% &  / & 1.731 \\
\noalign{\smallskip}\hline\noalign{\smallskip}
\end{tabular}
\begin{tablenotes}
      \small
      \item [*] ``H-TOP-1'' represents the TOP-1 of LLMs on the HumanEval~\cite{chen2021codex}.
    \end{tablenotes}
\end{threeparttable}
}
\end{table}

\vspace{-0.5cm}

\begin{table}[H]
\caption{AVG-5 results of 5 best-performing LLMs across 12 tiers.}
\label{tab:llm-difficuty}
\centering
\resizebox{.8\columnwidth}{!}{
\begin{tabular}{ccccccc}
\hline\noalign{\smallskip}
Tier & GPT-4 & GPT-3.5 & Claude & Bard & CodeLlama  \\
\noalign{\smallskip}\hline\noalign{\smallskip}
T1 & \cellcolor{gray!30}0.946 & \cellcolor{gray!15}0.879 & 0.715 & 0.577  & 0.262\\
T2 & \cellcolor{gray!30}0.885 & \cellcolor{gray!15}0.615 & 0.198 & 0.158  & 0.083\\
T3 & \cellcolor{gray!30}0.706 & \cellcolor{gray!15}0.576 & 0.267 & 0.190 & 0.051\\
T4 & \cellcolor{gray!30}0.664 & \cellcolor{gray!15}0.576 & 0.293 & 0.298  & 0.144\\
T5 & \cellcolor{gray!30}0.169 & \cellcolor{gray!15}0.096 & 0.012 & 0.012 & 0.000\\
T6 & \cellcolor{gray!30}0.493 & \cellcolor{gray!15}0.346 & 0.037 & 0.051  & 0.052\\
T7 & \cellcolor{gray!30}0.504 & \cellcolor{gray!15}0.363 & 0.113 & 0.061  & 0.060\\
T8 & \cellcolor{gray!30}0.285 & \cellcolor{gray!15}0.192 & 0.114 & 0.111  & 0.054\\
T9 & \cellcolor{gray!30}0.139 & \cellcolor{gray!15}0.139 & 0.031 & 0.018  & 0.000\\
T10 & \cellcolor{gray!30}0.186 & \cellcolor{gray!15}0.070 & 0.040 & 0.042  & 0.000\\
T11 & \cellcolor{gray!30}0.217 & \cellcolor{gray!15}0.130 & 0.026 & 0.026  & 0.000\\
T12 & \cellcolor{gray!30}0.000 & \cellcolor{gray!15}0.000 & 0.000 & 0.000  & 0.000\\
\noalign{\smallskip}\hline
\end{tabular}
}
\end{table}

\noindent\textbf{[Experiment Result]:}
Table \ref{tab:2} presents the repair performance of 12 LLMs, comprising 4 closed-source LLMs first, followed by 8 open-source LLMs. The table shows that the closed-source LLMs exhibit superior performance compared to the open-source LLMs across multiple evaluation metrics, including TOP-5, AVG-5, and RPSR. Regarding correctness metrics, TOP-5 and AVG-5, GPT-4 demonstrates the highest performance, achieving a TOP-5 of 66.5\% and an AVG-5 of 52.0\%, closely followed by GPT-3.5. The gap in performance between Claude/Bard and GPT-3.5/4 is notably significant within the TOP-5 and AVG-5 metrics. Among the open-source LLMs, CodeLlama and StarChat emerge as the best-performing for correctness, indicated by TOP-5 and AVG-5. In terms of AVG-5, CodeLlama achieves 6.8\%, while StarChat achieves 5.7\%. Except for CodeLlama and StarChat, Vicuna-13B exhibits the highest performance in terms of correctness, achieving merely a TOP of 1.9\% and an AVG-5 of 0.6\%. This result suggests that other open-source LLMs are ineffective for program repair tasks. Furthermore, among all the LLMs with AVG-5 scores exceeding 1\%, CodeLlama stands out with the lowest RPSR, indicating that CodeLlama excels in generating precise code patches. DrRepair, having only successfully repaired two incorrect codes, showcases the limited repair capabilities of traditional ITS techniques on \dataset compared to most selected LLMs.

The table also presents the TOP-1 results of LLMs on the HumanEval benchmark, denoted as H-TOP-1. Open-source LLMs, such as CodeLlama, CodeGen-multi-16B, StarChat, and CodeT5p, demonstrate H-TOP-1 results that are closely aligned with closed-source LLMs. Notably, the H-TOP-1 result of CodeLlama is nearly on par with that of GPT-3.5. However, when evaluated on \dataset, the AVG-5~(\ie the average TOP-1) results of open-source LLMs are considerably inferior to those of closed-source LLMs, which indicates considerable potential for improving open-source LLMs in practical repair scenarios.

\begin{figure}[H]
\centering
\includegraphics[width=.7\columnwidth]{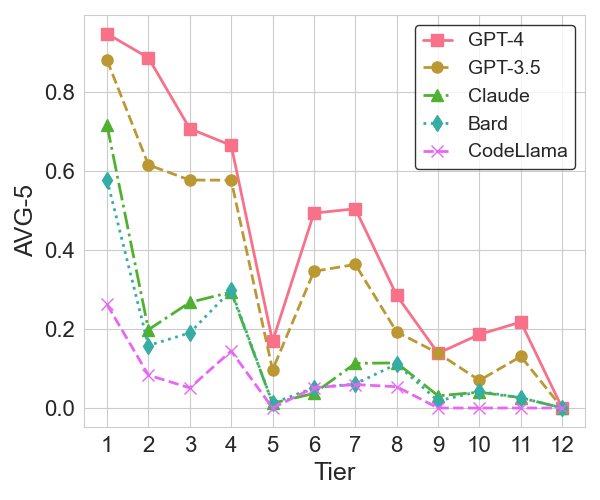}
\caption{AVG-5 trend of the 5 best-performing LLMs.}
\label{fig:llmlevel}
\end{figure}
\vspace{-0.5cm}
\begin{figure}[H]
\centering
\includegraphics[width=.8\columnwidth]{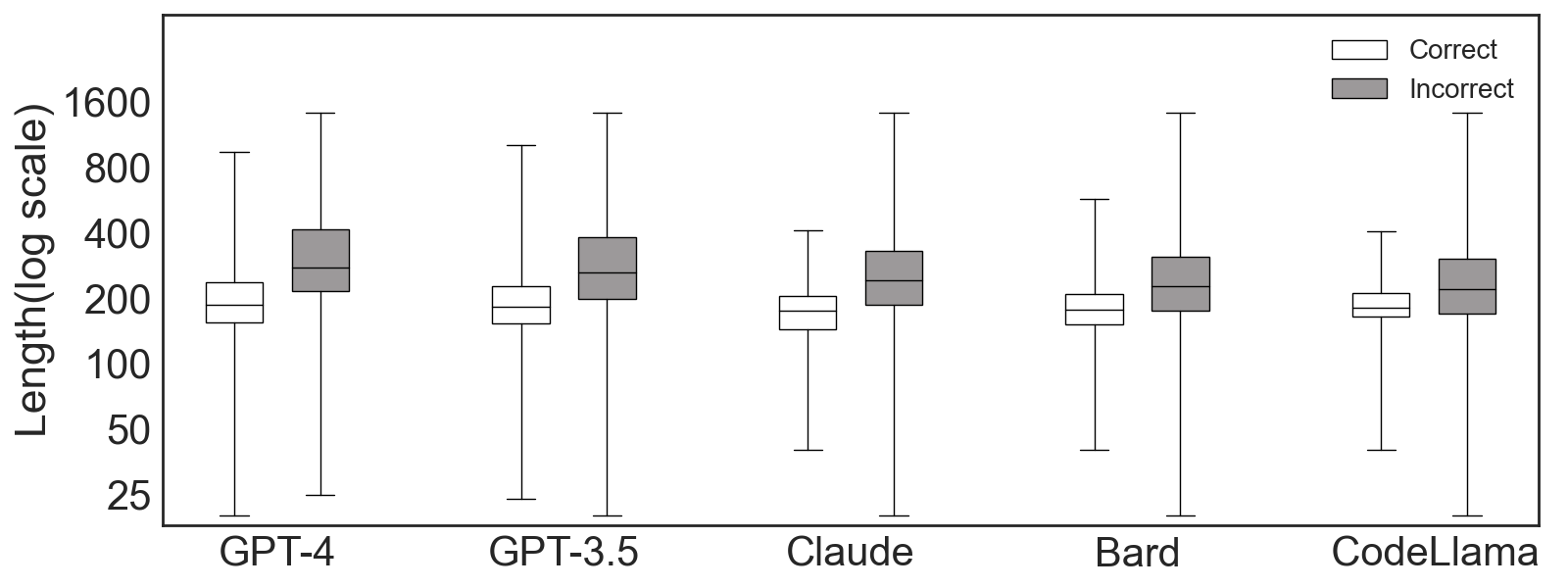}
\caption{Distributions of code lengths for 5 best-performing LLMs between correct and incorrect repair cases.}
\label{fig:prompt_len_baseline}
\end{figure}

We select 5 best-performing LLMs and compute their AVG-5 across 12 tiers, as illustrated in Table \ref{tab:llm-difficuty}. The table shows that GPT-4 and GPT-3.5 consistently demonstrate superior performance across all tiers. Notably, GPT-4 achieves an impressive AVG-5 of 94.6\% for tier T1, and it gradually declines with increasing tier, except tier T5. The unexpected low AVG-5 results for tier T5 are addressed in Section \ref{discussion}. Figure \ref{fig:llmlevel} emphasizes the descending trend of AVG-5 for each LLM across the 12 tiers.

Figure \ref{fig:prompt_len_baseline} illustrates the distributions of code lengths for correct and incorrect predictions generated by the 5 selected LLMs. It is observed that correct predictions~(present in white) tend to have shorter code lengths compared to incorrect predictions~(present in grey) across all the selected LLMs. Furthermore, it can be observed that longer incorrect codes adversely impact LLMs' repair capabilities, which aligns with the prior research~\cite{tian2023chatgpt}. Statistical significance between the correct and incorrect distributions was confirmed through the Mann-Whitney-Wilcoxon~(MWW)~\cite{mann1947test} test, supporting that correct predictions tend to have shorter code lengths. As illustrated in the figure, GPT-4 and GPT-3.5 exhibit better capabilities to handle lengthy prompts.

\find{{\bf [RQ-1]} {\bf Findings:} (1) In comparative analysis based on the AVG-5 metric, closed-source LLMs such as GPT-4 and GPT-3.5 demonstrate a superior performance, registering AVG-5 of 52.0\% and 41.5\%, respectively, in contrast to a meager 6.8\% recorded by leading open-source LLMs. (2) Although CodeLlama exhibits commendable performance on the HumanEval benchmark with a TOP-1 score of 42.7\%, its correctness significantly declines when evaluated on \dataset, trailing GPT-3.5 by a substantial AVG-5 margin of 34.7\%. This observation suggests that relying on small-scale benchmarks such as HumanEval may not accurately assess an LLM's repair capabilities. (3) Statistical analyses using box plots reveal that elongated code adversely impacts the repair performance of LLMs. {\bf Insights:} (1) The observed disparities in repair performance between closed-source and open-source LLMs highlight the imperative for the research community to redouble efforts to advance open-source LLMs. (2) Given the marked difference in performance metrics between \dataset and HumanEval, it is advisable for the research community to assess the repair capabilities of LLMs using large-scale, uncrawled benchmarks.}

\subsection{Enhancements of Augmented Information}
\noindent\textbf{[Experiment Goal]:}
We aim to explore the enhancements of augmented information on the repair capabilities of LLMs.

\noindent\textbf{[Experiment Design]:}
This experiment leverages 5 LLMs based on their outstanding performance in RQ-1, including GPT-4, GPT-3.5, Claude, and CodeLlama. The repair capabilities of LLMs are evaluated on \dataset. This experiment evaluates three levels of augmented information: general hints, specific guidance, and the automated execution result. Each of these three levels contains three types of information: solution description, tutor guidance, and failing test cases, as described in Section \ref{subsec:prompts}. The prompts are structured in the format described in Figure \ref{fig:prompts}. Seven different combinations of these three types of information are evaluated in this experiment, as detailed in the first column of Table \ref{tab:factors}. In the following paragraphs, we denote the combination of all three types of augmented information as \textbf{T\&S\&F}. Different types of augmented information are segmented into separate conversational entries within the T\&S\&F prompts to mitigate the adverse impacts of lengthy prompts. The repair capabilities of various combinations of augmented information are assessed in terms of two metrics: AVG-5 represents correctness, and RPSR represents patch precision. In our analysis of the enhancements of augmented information on the repair capabilities of LLMs across different tiers, we calculate AVG-5 for both the GPT-4 and GPT-3.5 across the 12 tiers for each combination of information. Subsequently, we analyze the corresponding performance enhancements and trends on tiers.

\begin{figure}[H]
\centering
\includegraphics[width=.7\columnwidth]{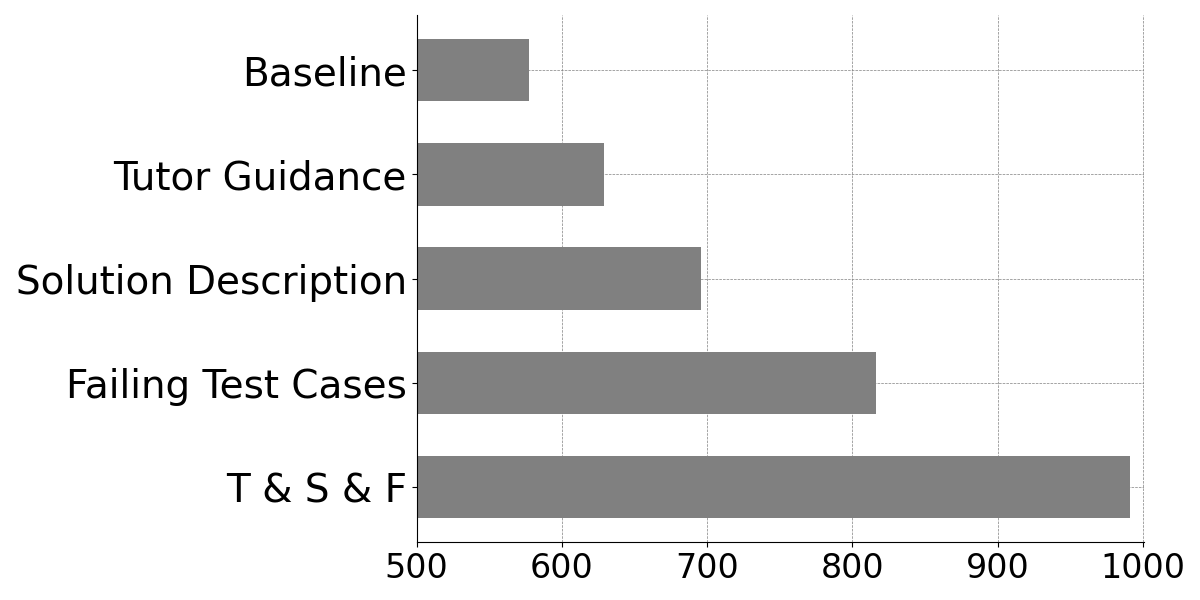}
\caption{Average tokens of prompt types with different augmented information.}
\label{fig:prompt_len_avg}
\end{figure}

Lengthy prompts have been found to have a detrimental influence on the program repair of LLMs~\cite{cao2023study,tian2023chatgpt,kemker2018measuring}. We calculate the length of input prompts with various combinations of information within \dataset, as illustrated in Figure \ref{fig:prompt_len_avg}. The extended length of prompts for failing test cases in the \dataset dataset is attributable to several factors. Unlike unit tests in other benchmarks, these test cases encompass comprehensive inputs and outputs, incorporating more data. Additionally, we present all failing test cases of the incorrect code within a single prompt. To reduce the negative impacts of lengthy prompts, we segment different types of information into separate conversational entries in this experiment.

\noindent\textbf{[Experiment Result]:}
Table \ref{tab:factors} presents the AVG-5 and RPSR results of the 5 selected LLMs. The table illustrates that among the three types of augmented information, tutor guidance emerges as the most effective in enhancing repair performance across all LLMs, as indicated by both AVG-5 and RPSR. Conversely, failing test cases negatively affect the repair performance of GPT-4, Claude, and Bard, as evidenced by both the AVG-5 and RPSR metrics. This deterioration in performance may be attributed to the lengthy prompt problem of failing test cases, as illustrated in Figure \ref{fig:prompt_len_avg}, the same as previous researchers have found~\cite{cao2023study,tian2023chatgpt,kemker2018measuring}. Meanwhile, it has been demonstrated that employing LLMs solely with failing test cases in repair tasks, like ChatRepair~\cite{xia2023conversation}, can not always surpass corresponding baselines on the \dataset when leveraging Claude, Bard, and CodeLlama.

For cases combining two types of augmented information, the combination of tutor guidance and solution description yields the most significant AVG-5 enhancements across all LLMs except Bard. Meanwhile, combining tutor guidance and failing test cases outperforms other combinations across all LLMs. The table shows that LLMs do not always attain the highest performance enhancements by T\&S\&F among all 7 combinations. For example, Bard, Claude, and CodeLlama do not achieve the best AVG-5 results. CodeLlama, in particular, produces a gap of 8.0\% compared to solely providing tutor guidance. All the selected LLMs provided with T\&S\&F do not achieve the best RPSR results compared to other combinations. These findings suggest that the issues of limited attention of LLMs persist even when all three types of augmented information are provided separately.

For any combination of augmented information, GPT-4 demonstrates a significant lead in the AVG-5 metric. The AVG-5 of other LLMs are ranked in descending order, with GPT-3.5, Claude, Bard, and CodeLlama following. Notably, CodeLlama exhibits the best RPSR result among all the selected LLMs. This finding further proves that CodeLlama achieves a higher patch precision, as discovered in RQ-1.

\begin{figure}[H]
    \centering
    \begin{subfigure}{0.48\columnwidth}
        \includegraphics[width=\columnwidth]{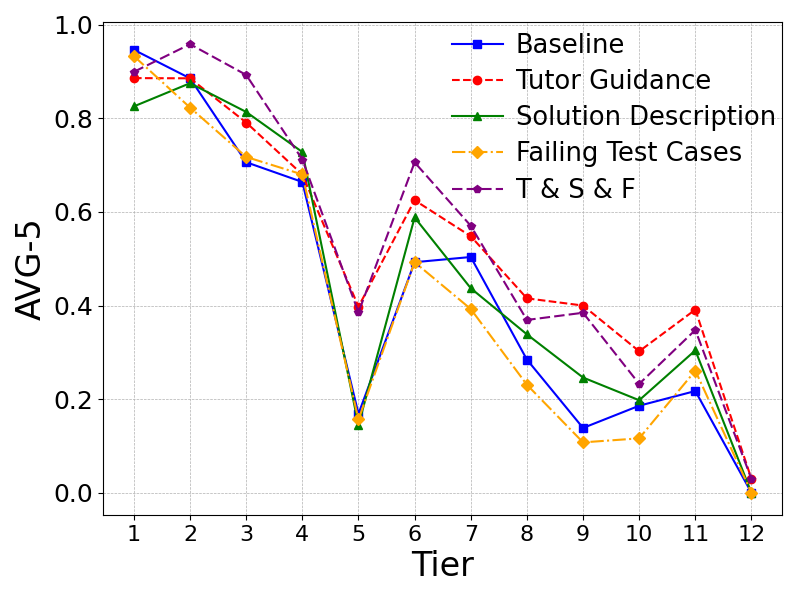}
        \caption{GPT-4}
        \label{fig:gpt4level}
    \end{subfigure}
    \begin{subfigure}{0.48\columnwidth}
        \includegraphics[width=\columnwidth]{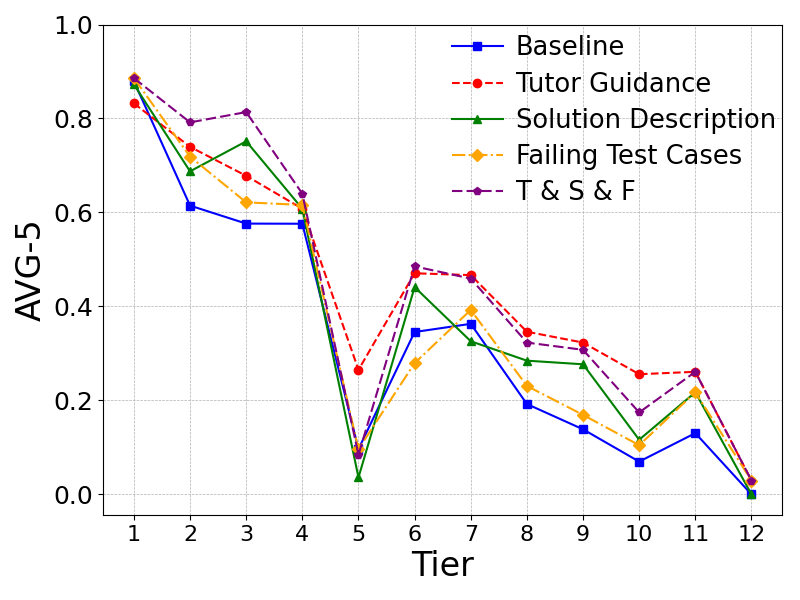}
        \caption{GPT-3.5}
        \label{fig:gpt3.5level}
    \end{subfigure}
    \caption{AVG-5 trend of (a)~GPT-4 and (b)~GPT-3.5 provided with three types of augmented information across 12 tiers.}
    \label{fig:gptlevel}
\end{figure}

The AVG-5 results for GPT-4 and GPT-3.5 across 12 tiers are illustrated in Figure \ref{fig:gpt4level} and Figure \ref{fig:gpt3.5level}. Notably, for GPT-4 and GPT-3.5, the AVG-5 results for tier T5 are significantly lower than the baseline when only the solution description is provided. This finding implies potential issues with the solution description of programming problems in tier T5, further analyzed in Section 5. Furthermore, the AVG-5 for each successive tier exhibits a decreasing trend, except for tier T5.

\vspace{-0.2cm}
\vspace{-0.5cm}
\begin{table}[H]
\caption{AVG-5 and RPSR results of 5 LLMs provided with various combinations of augmented information.}
\centering
\label{tab:factors}
\setlength{\tabcolsep}{2pt}
\resizebox{\columnwidth}{!}{
\begin{threeparttable}
\begin{tabular}{c*{12}{c}}
\toprule
Combination & \multicolumn{2}{c}{GPT-4} & \multicolumn{2}{c}{GPT-3.5} & \multicolumn{2}{c}{Claude} & \multicolumn{2}{c}{Bard} & \multicolumn{2}{c}{CodeLlama}\\
\cmidrule(lr){2-3}
\cmidrule(lr){4-5}
\cmidrule(lr){6-7}
\cmidrule(lr){8-9}
\cmidrule(lr){10-11}
& AVG-5  & RPSR & AVG-5  & RPSR & AVG-5  & RPSR & AVG-5  & RPSR & AVG-5  & RPSR \\
\midrule
Baseline & 52.0\%  & 3.748 & 41.5\% & 5.072 & 20.3\%  & 4.083 & 16.8\%  & 4.728 & 6.8\% & 3.635 \\
\midrule
Tutor Guidance~(T) & 61.4\%  & 2.210  & 50.9\%  & 2.950 & 27.1\%  & 1.834 & \cellcolor{green!20} 26.7\%  & 1.939 & \cellcolor{green!20}16.8\% & 2.042 \\
Solution Description~(S) & 55.6\%  & 4.069 & 47.4\%  & 5.638 & 21.3\%  & 4.485 & \cellcolor{gray!15} 16.7\% & 4.393 & 7.7\%  & 3.631 \\
Failing Test Cases~(F) & \cellcolor{gray!15}49.7\% & 4.366  & 42.2\%  & 6.360 & \cellcolor{gray!15} 19.4\%  & 5.285 & \cellcolor{gray!15} 15.1\%  & 5.112 & 7.5\% & 4.005 \\
\midrule
T \& S & 61.9\%  & 2.947  & 51.3\% &3.069  & \cellcolor{green!20} 27.6\% & 2.568 & 19.3\%  & 2.497 & 14.8\% & 3.011 \\
T \& F & 59.7\% & 2.736  & 51.3\%  & 3.490 &  27.5\%  & 2.101 & 26.6\%  & 2.231 & 16.3\% & 1.980 \\
S \& F & 53.2\%  & 4.258 & 47.0\% & 6.556 &  24.9\%  & 4.667 & \cellcolor{gray!15} 15.9\%  & 5.098 & 7.9\% & 5.383 \\
\midrule
T \& S \& F & \cellcolor{green!20}62.3\%  & 2.778  & \cellcolor{green!20}52.4\% & 5.201 & 27.6\%  & 3.213 & 24.4\%  & 4.235 & 8.8\% & 3.376\\
\bottomrule
\end{tabular}
\begin{tablenotes}
      \small
      \item Green cells show the best-performing combination. Gray cells are below the baseline.
    \end{tablenotes}
\end{threeparttable}
}
\end{table}
\vspace{-0.7cm}
\vspace{-0.3cm}

\find{{\bf [RQ-2]} {\bf Findings:} (1) Among the three types of augmented information, tutor guidance notably outperforms the others in enhancing the repair capabilities of LLMs, showing a significant increase of over 6.8\% in the AVG-5 metric across all evaluated LLMs. (2) Conversely, the simple integration of multiple types of augmented information does not consistently produce benefits over the baseline. For instance, in terms of Bard, incorporating both solution description and failing test cases results in a 0.9\% AVG-5 decrease relative to the baseline. {\bf Insights:} Human guidance information improves the repair capabilities of LLMs significantly; it is valuable for future researchers to explore assisting rather than replacing humans in the programming education scenarios.}

\subsection{Conversational Program Repair}
\noindent\textbf{[Experiment Goal]:}
We introduce an LLM-based conversational semi-automatic repair framework \tool and investigate its repair capabilities using augmented information through automated evaluation.

\vspace{-0.2cm}
\begin{figure}[H]
\centering
\includegraphics[width=\columnwidth]{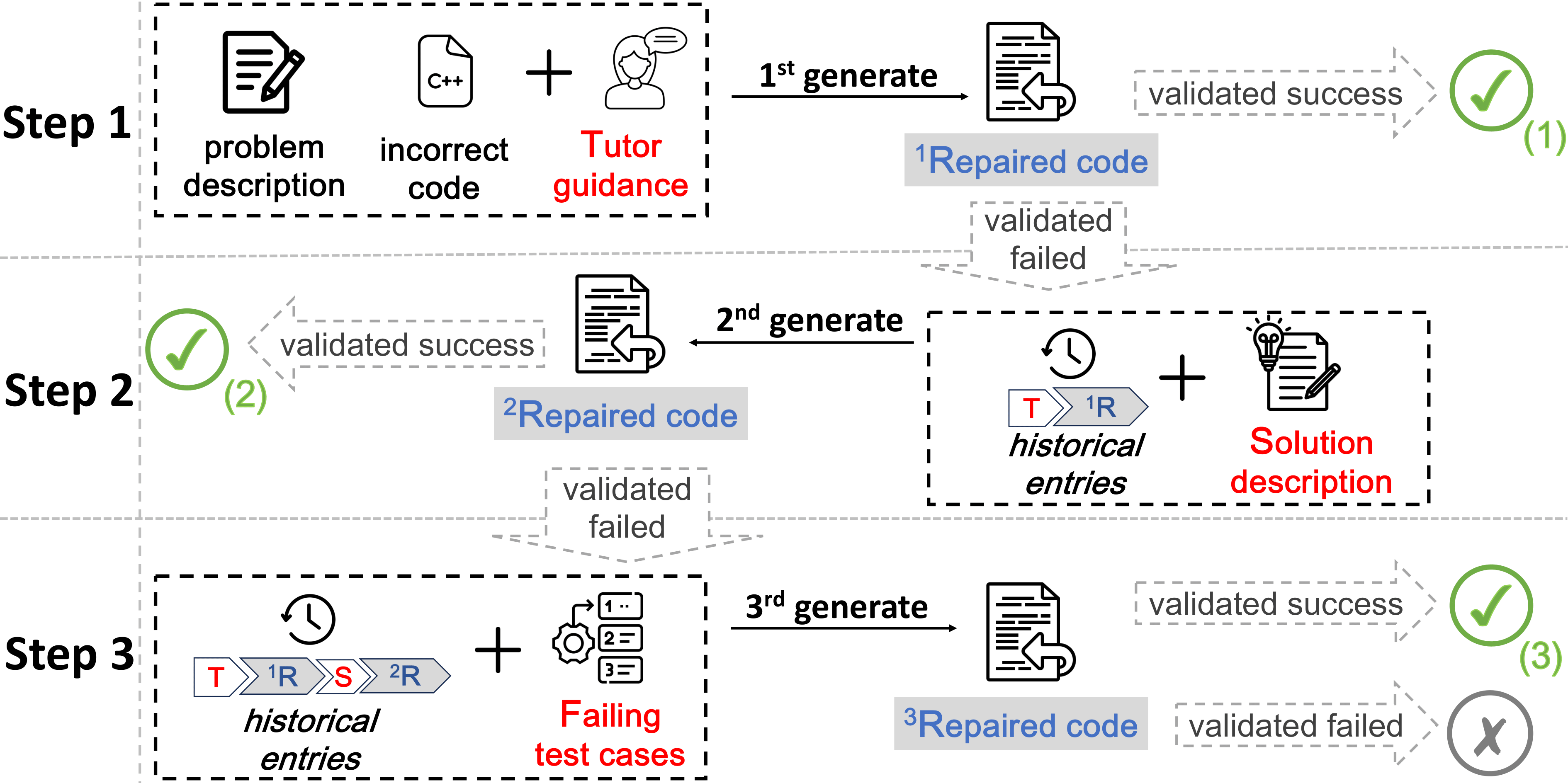}
\caption{Overview of \tool.}
\label{fig:CREF}
\end{figure}
\vspace{-0.5cm}

\noindent\textbf{[Experiment Design]:}
This experiment employs five top-performing LLMs identified in RQ-1, including GPT-4, GPT-3.5, Claude, Bard, and CodeLlama, which are consistent with the methodology used in RQ-2. Utilizing LLMs with various types of augmented information often results in lengthy prompts, potentially leading to issues of limited attention. To counteract the detrimental effects of lengthy prompts, we introduce a repair strategy termed \othertool. This strategy partitions the repair process into three phases: tutor guidance, solution articulation, and the presentation of failing test cases. The \othertool initiates three distinct dialog sessions for each incorrect code, thereby mitigating the detrimental effects of lengthy prompts. In programming education scenarios, it is crucial to accelerate the process of repairing students' debugging requests~\cite{lee2023learning}. Tutor guidance, due to its considerable impact on enhancing performance~(shown in Table \ref{tab:factors}), is prioritized as the first in our approach. Subsequently, LLMs are furnished with solution descriptions, ranking second in performance improvement. Lastly, LLMs are provided with failing test cases, which yield the least performance enhancement related to lengthy prompts.

Although \othertool mitigates the adverse effects of lengthy prompts, it discards historical conversation entries. However, the historical entries could potentially aid LLMs in avoiding the same errors in subsequent generations~\cite{xia2023conversation} and fully utilizing conversational capabilities, which is similar to the idea of chain-of-thought~(CoT)~\cite{wei2022chain}. To unlock the potential conversational capabilities of LLMs, we propose a repair framework named \tool. Figure \ref{fig:CREF} illustrates the overview of \tool. In \tool, there is only one dialog session instead of three dialog sessions compared to \othertool. The \tool follows three steps: (1) Providing LLMs with programming problem description, incorrect code with corresponding tutor guidance, and generates repaired code. If the repaired code is validated successfully, then the process ends. (2) Providing LLMs with all historical entries and solution descriptions to generate a repaired code. If the repaired code is validated successfully, the process ends. (3) Providing LLMs with all historical entries with failing test cases of the last repaired code to generate the repaired code. The success of the repair process is determined by whether or not there is a correct repaired code so far.

We evaluate the repair performance of two proposed methodologies, \othertool and \tool, utilizing 5 selected LLMs on the \dataset, in terms of AVG-5 and RPSR. We compare these results with the baseline and T\&S\&F. We further evaluate the AVG-5 results of \tool utilizing GPT-4 and GPT-3.5 across 12 tiers to analyze trends and enhancements compared to the baseline. 

To illustrate \tool's scalability, we compared its repair capabilities with baseline, T\&S\&F, and \othertool, using GPT-4 and GPT-3.5, on \textbf{\datasetextend}. \datasetextend resembles \dataset but includes a broader selection of 2,464 incorrect code submissions. These submissions, which come with tutor guidance, are created by 786 students across 45 distinct programming problems, consistently classified into 12 tiers. \datasetextend integrates all 35 problems from \dataset, incorporating its original 1,239 instances alongside additional new samples. On average, each problem in \datasetextend includes 6.98 pairs of well-designed test cases. Notably, while \dataset features an average of 1.89 functions per incorrect code, \datasetextend showcases a similar complexity with an average of 1.81 functions. \datasetextend does not guarantee the inclusion of codes corrected by the students, leading us to omit the RPSR metric on \datasetextend. Due to the company's commercial considerations, \datasetextend will remain proprietary, but we will update \dataset with more samples to support the program repair research community.

\vspace{-0.2cm}
\begin{table}[H]
\caption{AVG-5 and RPSR results of T\&S\&F, \othertool, and \tool compared to the baseline.}
\centering
\label{tab:chat_all}
\setlength{\tabcolsep}{2pt}
\resizebox{\columnwidth}{!}{
\begin{threeparttable}
\begin{tabular}{c*{8}{c}}
\toprule
Model & \multicolumn{2}{c}{Baseline} & \multicolumn{2}{c}{T\&S\&F} & \multicolumn{2}{c}{MultiRegenerate} & \multicolumn{2}{c}{\tool} \\
\cmidrule(lr){2-3}
\cmidrule(lr){4-5}
\cmidrule(lr){6-7}
\cmidrule(lr){8-9}
& AVG-5  & RPSR & AVG-5~(↑) & RPSR  & AVG-5~(↑)  & RPSR & AVG-5~(↑) & RPSR \\
\midrule
GPT-4  & 52.0\%  & 3.748  & 62.3\%~(+10.3\%)  & 2.778 & 71.5\%~(+19.5\%) & 2.815 & \cellcolor{green!20}76.6\%~(+24.6\%) & 2.691 \\
GPT-3.5  & 41.5\%  & 5.072 & 52.5\%~(+11.0\%)  & 5.201  & 62.6\%~(+21.1\%) & 3.438 & \cellcolor{green!20}63.8\%~(+22.3\%) & 3.454 \\
Claude  & 20.3\%  & 4.083  & 27.6\%~(+~7.3\%) & 3.213 & 38.9\%~(+18.6\%) & 2.782 & \cellcolor{green!20}42.3\%~(+22.0\%) & 2.768 \\
Bard  & 16.8\%  & 4.728  & 24.4\%~(+~7.6\%) & 4.235 & 35.2\%~(+14.9\%) & 2.562  & \cellcolor{green!20}35.4\%~(+18.6\%) & 2.618 \\
CodeLlama  & 6.8\% & 3.635 & ~8.8\%~(+~2.0\%) & 3.376 & 22.5\%~(+15.7\%) & 2.781 & \cellcolor{green!20}24.0\%~(+17.2\%) & 2.880 \\
\bottomrule
\end{tabular}
\begin{tablenotes}
      \small
      \item A smaller RPSR is better, meaning the patch is more precise.
    \end{tablenotes}
\end{threeparttable}
}
\end{table}
\vspace{-0.3cm}

\noindent\textbf{[Experiment Result]:}
Regarding AVG-5, \tool exhibits superior lead performance, outperforming the baseline and T\&S\&F by margins of 24.6\%~(GPT-4) and 15.2\%~(CodeLlama), respectively, and surpassing \othertool. Although \othertool achieves satisfactory repair performance, initiating three distinct dialog sessions for each repair requires more computational resources in practical engineering scenarios. Regarding RPSR, both \othertool and \tool exhibit significant improvements compared to the baseline and T\&S\&F. A lower RPSR value indicates greater precision of patches. Compared to \othertool, which modifies the initial incorrect code in a single iteration, \tool interactively repairs the initial incorrect code over multiple rounds of dialogue. Despite this, \tool maintains a comparable RPSR to \othertool while achieving a significant lead in AVG-5.

The AVG-5 trends of \tool and baseline across 12 tiers when utilizing GPT-4 and GPT-3.5 are illustrated in Figure \ref{fig:gpt4level} and Figure \ref{fig:gpt3.5level}, respectively. \tool exhibits a significant improvement over the baseline across all 12 tiers. Furthermore, the AVG-5 results of \tool exhibit a decreasing trend except for tier T5.

\textbf{Results on Extend Benchmark:} We validate the generalization of \tool's repair capability on an extensive benchmark \datasetextend. On \datasetextend benchmark, when prompt LLMs with the baseline, the AVG-5 result of GPT-4 is 49.3\%, and the AVG-5 result of GPT-3.5 is 39.6\%. Both are close to the corresponding results in RQ-1. When providing LLMs with T\&S\&F, the AVG-5 of GPT-4 reaches 60.9\%, while the AVG-5 of GPT-3.5 reaches 50.1\%, which is an improvement of 11.6\% and 10.5\% compared to the baseline respectively, and is similar to the results in RQ-2. When leveraging LLMs with \othertool strategy, the AVG-5 of GPT-4 reaches 68.3\%, while the AVG-5 of GPT-3.5 reaches 60.3\%. AVG-5 of \tool utilizing GPT-4 reaches \textbf{74.4\%} while AVG-5 of \tool utilizing GPT-3.5 reaches 61.9\%. For GPT-4 and GPT-3.5, the AVG-5 of \tool compared to \othertool is higher by \textbf{6.1\%} and 1.6\%, respectively, and this gap is higher than that of the results on \dataset. These results demonstrate the scalability of our findings.

\find{{\bf [RQ-3]} {\bf Findings:} (1) In assessing the selected five LLMs, \tool shows marked improvements across both the AVG-5 and RPSR metrics when compared to the baseline and the T\&S\&F. Notably, every LLM in this experiment exhibits an increase in AVG-5 performance by a minimum of 17.2\% over the baseline, with GPT-4 outperforming all others by achieving an impressive 24.6\% improvement. (2) When compared with \othertool, \tool maintains a similar RPSR while securing a higher AVG-5. These results highlight the advantage of integrating historical conversational entries to enhance the repair capabilities of LLMs. (3) Utilizing GPT-4 and GPT-3.5, the AVG-5 results of the baseline, T\&S\&F, \othertool, and \tool on the \datasetextend support the generalizability of our earlier findings. {\bf Insights:} The performance of \tool up to 76.6\% of AVG-5 suggests that \tool is a useful semi-automatic repair tool in education scenarios for assisting programming tutors, and future research could further explore improving patch precision for the students.}

\vspace{-0.5cm}
\begin{figure}[H]
\centering
\includegraphics[width=.7\columnwidth]{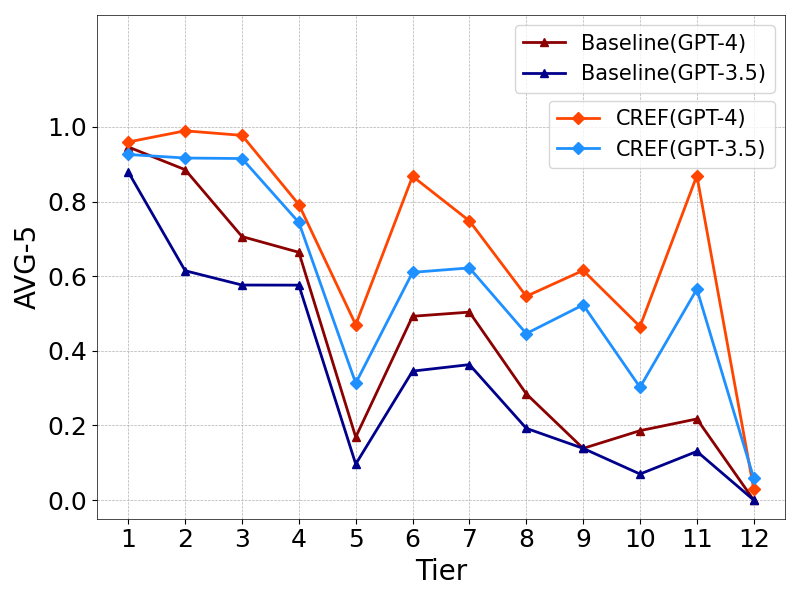}
\caption{AVG-5 trend of \tool utilizing GPT-4/GPT-3.5.}
\label{fig:Convlevel}
\end{figure}
\vspace{-0.5cm}
\section{Discussion}
\label{discussion}
\subsection{Investigation of Outlier Data}
We discuss the generally less-than-expected AVG-5 results of LLMs for Tier 5 to investigate further the factors affecting the repair performance. Tier 5 incorporates two programming problems: \textit{Onion Girl's Flying Chess} and \textit{Artificial Intelligence}. A subsequent investigation reveals inherent ambiguities in the statements of both programming problems and in the solution description of \textit{Onion Girl's Flying Chess}. To address this, we implemented targeted modifications to resolve these ambiguities and recalculated the AVG-5 scores for GPT-3.5 and GPT-4 on tier T5 employing the framework \tool. Following these adjustments, the AVG-5 score for GPT-4 increases to 76.7\%, a gain of 29.7\%, whereas GPT-3.5 achieves an AVG-5 of 59.9\%, reflecting an improvement of 28.6\%. These findings suggest that the clarity of the contextual information provided to LLMs significantly impacts their repair capabilities.

\subsection{Industrial Application}
On the company's programming education platform, students learn to code in C++ by working on given assignments. Novice students often require guidance to repair incorrect code and yield a correct solution. Tutors hired by the company help students debug their code, but this process can be labor-intensive and usually involves long response times. To address these challenges, we introduced \tool assisting tutors on a programming education platform to address students' debugging requests. First, tutors only need to provide preliminary guidance instead of debugging solutions, as shown in Figure \ref{fig:def_reply_eg}. \tool can automatically complete the repair process in most cases. Subsequently, tutors review \tool’s generation to formulate a response to students, ensuring precise and effective debugging assistance. Experimental results on \dataset, where the average tutor guidance consists of just 38.25 words, show that \tool utilizing such preliminary feedback significantly increases the success rate of automated program repairs.

The two-month deployment of \tool has demonstrated significant reductions in labor costs. The reduction in average response time for a debugging request, from 26.7 minutes to 7.7 minutes, is attributed to the utilization of \tool. \tool has shifted the tutors' role towards providing preliminary guidance rather than full debugging solutions. The reduction in labor costs by 71.2\% indicates that the same labor cost can address 3.5 times more debugging requirements from students, making it a cost-effective tutoring solution.

The success of \tool in educational contexts suggests its potential for broader applications, such as code reviews, where it utilizes reviewer feedback to refine codes, offering developers enhanced solutions automatically. This application of \tool directly contributes to faster code reviews, highlighting its adaptability to various code-related tasks.

\section{Threats to Validity}
\label{threats}

\noindent
{\sc \em Threats to External Validity.} The generalizability of our findings is influenced by two factors: the scale of benchmarks and the risk of data leakage. Consequently, we mitigate these threats by employing a large-scale uncrawled benchmark \dataset, comprising 1,239 instances. The selection of benchmarks not only aims to cover a wide array of bugs but also mitigate the risk of data leakage, thereby enhancing both the generalizability and reliability of our study.

\noindent
{\sc \em Threats to Internal Validity.} The randomness of LLM outputs influences the reliability of our experimental results. To mitigate this influence, we have adopted a five-fold repetition strategy in our experiments, allowing for a more reliable calculation of the performance metrics TOP-5 and AVG-5. Furthermore, we have standardized the key parameters including $temperature$ and $top\_p$ for all LLMs in our experiments to mitigate the potential influence of parameter variations on the reliability of our comparative analyses. Additionally, automated scripts were employed to extract code from LLM-generated non-compilable natural language explanations when provided with incorrect codes and contextual information to extract compilable codes, ensuring our analysis is based on usable data.

\noindent
{\sc \em Threats to Construct Validity.} The selection of evaluation metrics influences the validity of our findings. To mitigate risks associated with metric selection and ensure accurate evaluation results, we selected metrics TOP-5, AVG-5, and the RPSR. These metrics are well-known for measuring the correctness and precision of LLMs. To address concerns that these metrics potentially may lead to biased conclusions, we evaluated \tool on extensive benchmark \datasetextend, which includes 2,464 samples. \datasetextend helps us ensure our findings are solid and can be applied to diverse coding bugs.

\section{Conclusion}
\label{conclusion}

In this paper, we introduce an extensive uncrawled benchmark \dataset consisting of 1,239 C++ defect codes and types of associated information. Utilizing \dataset, experiments are conducted to investigate the realistic repair performance of 12 prominent LLMs, and demonstrate the significant difference on HumanEval and \dataset. We investigate how augmented information improves the repair capabilities of best-performing LLMs. The experimental results show that tutor guidance improves the LLM repair performance the most, while failing test cases improve the least due to the lengthy prompt problem and even degrades the repair performance for Bard and Claude. To mitigate these negative impacts, we propose a strategy \othertool, to minimize the adverse effects of lengthy prompts, which repairs incorrect code through three distinct conversational sessions. Finally, we introduce a novel conversational semi-automatic repair approach \tool, which is engineered to optimize the utilization of augmented information and conversational capabilities of LLMs. Experimental results indicate that the repair performance of \tool is a significant lead in AVG-5 and RPSR metrics compared to the baseline and T\&S\&F, and yields superior AVG-5 and comparable RPSR results compared to \othertool. These results indicate that incorporating historical failing repairs can significantly enhance repair capabilities in LLMs by fully exploiting their conversational potential. \tool acts as an assisting tool through a three-step process: (1) tutors provide preliminary guidance of the student's buggy code, (2) based on this guidance, \tool automatically performs code debugging, and (3) tutors leverage \tool’s generation to reply to students. This approach has cut response times by 71.2\% and reduced costs by 69.9\%, improving the tutoring process and student learning experiences. The success of \tool highlights its potential for broader uses, such as augmenting code review processes by automatically adjusting codes based on reviewers' comments, suggesting future applications for enhancing efficiency in coding-related tasks.

While ChatGPT shows strong performance, its training and inference require substantial computation due to huge model parameters. Future research could focus on optimizing other open-source LLMs with augmented information and conversational capabilities to reduce computational costs, especially for enhancing programming guidance in educational scenarios.


\section*{Open Science}
The artifact of this study is publicly available at {\bf\url{https://github.com/buaabarty/CREF}}.

\bibliographystyle{ACM-Reference-Format}
\bibliography{sample-base}

\end{document}